\documentclass[]{revtex4-1}

\usepackage[T1,T2A]{fontenc}
\usepackage[utf8]{inputenc}
\usepackage[english]{babel}

\usepackage{amsmath,amssymb}
\usepackage{graphicx}
\usepackage{comment} 
\usepackage{color}
\usepackage{ulem}

\DeclareMathOperator{\St}{St}
\newcommand{\tsc}{\tau_\mathrm{s}}

\renewcommand{\vec}[1]{\boldsymbol{\mathbf{#1}}}

\begin{document}

\title{Anomalous Conductivity and Anisotropic Transport of~Nonrelativistic Electrons in~Plasma with~Magnetostatic Weibel-Generated Turbulence}

\author{Nikolay~A.~Emelyanov}
\affiliation{Institute of Applied Physics of the Russian Academy of Sciences, Nizhny Novgorod, Russia}
\affiliation{Pulkovo Observatory of the Russian Academy of Sciences, Saint Petersburg, Russia}

\author{Mikhail~A.~Garasev}
\affiliation{Institute of Applied Physics of the Russian Academy of Sciences, Nizhny Novgorod, Russia}

\author{Aleksey~A.~Kuznetsov}
\affiliation{Institute of Applied Physics of the  Russian Academy of Sciences, Nizhny Novgorod, Russia}

\author{Anton~A.~Nechaev}
\affiliation{Institute of Applied Physics of the  Russian Academy of Sciences, Nizhny Novgorod, Russia}

\author{Evgenii~A.~Shirokov}
\email[]{eshirokov@ipfran.ru}
\affiliation{Institute of Applied Physics of the  Russian Academy of Sciences, Nizhny Novgorod, Russia}

\author{Vladimir~V.~Kocharovsky}
\affiliation{Institute of Applied Physics of the  Russian Academy of Sciences, Nizhny Novgorod, Russia}

\date{\today}

\begin{abstract}
The anisotropic diffusion of electrons with various rigidity and the anomalous conductivity of a collisionless plasma in the presence of Weibel-generated quasi-static turbulent and uniform external magnetic fields are examined. Using an original code based on the Boris algorithm, the electron diffusion coefficients and the longitudinal, transverse, and Hall mobility factors are determined for a representative set of plasma parameters. It is shown that these values and their anisotropy depend strongly on the electron temperature, external magnetic field, average level of magnetic turbulence, and its spectrum. The physical origin and expected limits of such dependencies are indicated. Applications of the results are discussed in the case of coronal plasma, where the anomalous resistivity prevails over the collisional one and can be responsible for the redistribution of large-scale currents in magnetic loops and small-scale currents in the regions of reconnection of magnetic-field lines.
\end{abstract}


\maketitle

\section{Introduction}

The finding of transport characteristics of a turbulent plasma, such as anomalous viscosity, heat- and electro- conductivity, is a fundamental problem that can shed light on the physical mechanisms of many phenomena in laboratory and cosmic plasmas (see, e.g., \citep{Rosenbluth1966, Biskamp1993, Krommes2002, Matthaeus2011, Karimabadi2013, Matthaeus2021, Vogman2025, Graham2025}).  For the considerably hot and rarefied plasmas, which are typical for cosmic environments like the magnetospheres of planets, solar and stellar winds or corona, the interaction of charged particles with turbulent fluctuations of the electric and magnetic fields has a more significant impact on the medium transport properties than interparticle collisions~\citep{Norman1978, Karimabadi2013, Bian2016, Yoon2017, Bian2018, Khotyaintsev2019, Yoon2019, Graham2022, Isliker2025}. 

No wonder that a huge amount of works has been devoted to this problem. Most of the authors focus on the interaction of particles with electrostatic fluctuations~\citep{  Rostoker1961, Davidson1972,  Akhiezer1975, Che2017, Khotyaintsev2019, Vogman2025, Graham2025, Yoo2025} (e.g., ion-acoustic or Langmuir ones), which is expected to prevail over their interaction with electromagnetic or quasi-magnetostatic turbulence in the case of nonrelativistic particles. Nevertheless, under some conditions, the anomalous transport properties of collisionless plasma can be determined by very-low-frequency electromagnetic or quasi-magnetostatic turbulence. It can appear in the absence of a strong external magnetic field due to several kinetic instabilities in a nonequilibrium plasma, i.g., the firehose or Weibel-type ones, according to a number of experiments~\citep{Carter2002, Ji2004, Ji2023}, numerical PIC simulations~\citep{Daughton2011, Tummel2020, Wang2021, Nechaev2022, Graham2025, Vicentin2025} and spacecraft observations in the Earth's magnetosphere~\citep{Ergun2020, Cozzani2021, Graham2022, Graham2025} and solar wind~\citep{Matthaeus2021, Lotekar2022, Howes2024}. 

The particle transport through magnetic turbulence is deeply studied in the theory of cosmic ray propagation, which has a long history~\citep{Medvedev2000, Casse2001, Plotnikov2011, Keenan2013, Shalchi2020, Zhang2024, Isliker2025, Kuhlen2025b, Kuhlen2025a} and can be traced back far into the past (see, e.g.,~\citep{Ginzburg1966, Jokipii1966, Jokipii1967, Forman1974, Giacalone1994}). Recently, the phenomenon of anisotropic diffusion of charged particles under magnetic turbulence has been extensively analyzed not only for cosmic rays~\citep{Shalchi2020, Zhang2024, Kuhlen2025b, Jokipii1966, Reichherzer2025}, but also in the solar atmosphere~\citep{Shalchi2009, Kontar2011, Bian2016, Emslie2018, emslieEffectsTurbulentElectrical2024,emslieEffectsTurbulentElectrical2025}, the solar wind~\citep{Jokipii1967, Pommois2001, Chhiber2021}, and the Earth's magnetotail~\citep{Lazarian2020, Lemoine2023}. However, in most cases, this analysis is given for the magnetohydrodynamics (MHD) turbulence and is limited to either ions or hot electrons, i.e., it does not consider the main part of plasma electrons, which are relatively cold. 

For what follows, note that the kinetic quasi-magnetostatic turbulence can be a dominant factor in the anomalous plasma conductivity. For example, it can be responsible for the enhanced ohmic heating of coronal and chromospheric plasma by a return electric current produced by the injected non-thermal electrons during the impulse phase of the solar flare, as discussed in \citep{Bian2016,emslieEffectsTurbulentElectrical2024,emslieEffectsTurbulentElectrical2025}. Also, the magnetic turbulence is inevitable in the reconnection phenomena leading to the fast energy release and rearrangement of the self-consistent current structures in active regions of the solar corona~\cite{Ji2007, Aschwanden2019, Arnold2021, Dahlin2022, Graham2025, Sen2025}. This rearrangement is related to an inhomogeneous anomalous conductivity and can initiate multiple flares, e.g., in the global current layers over the coronal loop apexes, as in the CSHKP model~\citep{Shibata2011, Benz2016, Gritsyk2021}, or in the system of small current sheets at the contact areas between sheared magnetic loops, as in the alternative approaches; see, e.g.,~\citep{Parker1983a, Shibata2011}. 

The contribution of the turbulent magnetic field to the effective electric resistivity of various particle species in plasma systems and its role in different astrophysical and laboratory phenomena are poorly investigated. In particular, there is no adequate analysis of the properties of the anomalous anisotropic mobility of the dominant fraction of electrons in cosmic plasma with kinetic magnetic turbulence and a homogeneous external magnetic field. To our knowledge, the most advanced yet illustrative phenomenological evaluation of the parallel electrical conductivity, accounting for magnetic turbulence, was carried out in the work~\citep{Bian2016}. It is based on the Lorentz plasma model with both collisional and turbulent scattering of electrons in the presence of a homogeneous background magnetic field and the transverse magnetostatic fluctuations with the Lorentz-type longitudinal spectrum.

In this article, we provide the numerical studies and analytical estimates of the anomalous electron conductivity and diffusion in a collisionless nonrelativistic plasma, which are determined by the interaction of particles with kinetic small-scale quasi-magnetostatic turbulence. Namely, the samples of this turbulence are taken as the results of the nonlinear evolution of the kinetic Weibel-type instability~\citep{Weibel1959, Hasegawa1975, Kocharovsky2016} for an initial-value problem in an anisotropic plasma with the bi-Maxwellian velocity distribution of electrons in the presence of a relatively weak external magnetic field. We verify the diffusion regime of particle motion and compute components of the electron mobility tensor using an original code based on the well-known Boris algorithm~\citep{Boris1970, Arber2015}. On the basis of typical examples, we analyze how the longitudinal and transverse diffusion coefficients depend on the electron rigidity, as well as how the mobility tensor components depend on plasma electron temperature, the external magnetic field, the average level of magnetic turbulence, and the width and anisotropy of its spectrum.

The article contains the following sections. The analytical relations between the diffusion coefficients and mobility-tensor components are presented in Section II. The original numerical code and main features of computer calculations are given in Section III. Typical examples of the diffusion coefficients as functions of the rigidity parameter, as well as the dependence of the anisotropic anomalous conductivity on the electron temperature and the ratio between the average turbulent magnetic field and the total field, are shown in Section IV. Section V contains final comments on the physical interpretation of the results, their application to the processes in coronal plasma and some open problems, as well as the general conclusions.

\section{Derivation of the Einstein Relations}
\label{sec:tensors}

Consider a nonrelativistic collisionless plasma in a static magnetic field $\vec{B} = \vec{B}_\mathrm{ext} + \vec{B}_\mathrm{turb}$, where $\vec{B}_\mathrm{ext}$ is an external uniform magnetic field and $\vec{B}_\mathrm{turb}$ is a turbulent component (random on small scales). For definiteness, in our simplistic model, we take the component $\vec{B}_\mathrm{turb}$ as a snapshot of the field produced by the Weibel-type instability. Although neglecting Coulomb collisions, we treat electron scattering in momentum space on magnetic field inhomogeneities as diffusion and account for it in the Boltzmann equation by using a collision term $\St[f]$:
\begin{equation}
    \frac{\partial f}{\partial t} + \vec{v} \nabla f + \frac{q}{m}\left(\vec{E}+\frac{\vec{v}}{c} \times \vec{B}_\mathrm{ext}\right) \nabla_{\vec{v}} f = \St[f] ,
\label{eq:kin}
\end{equation}
where $f(t,\vec{r},\vec{v})$ is the electron distribution function normalized to the number density $n = \int \! f \, d^3\vec{v}$, $\vec{E}$ is a weak external electric field applied to the plasma, $q = -e$ and $m$ are the electron charge and mass, and $c$ is the speed of light in a vacuum. Heavy ions are deemed immobile. We assume that the collision (scattering) operator is linear and
\begin{equation}
    \St[f_0] = 0 ,
	\
	\int \! \St[f_1] \,d\phi \,d\beta = 0 ,
\label{eq:StProp}
\end{equation}
where $f_0(t,\vec{r},v) = \int \! f \,d\phi \,d\beta$ is the velocity isotropic part of the distribution function and $f_1 = f - f_0$.
Here $\phi$ and $\beta = \cos\theta$ are the azimuthal angle and the cosine of the pitch angle $\theta$ in spherical coordinates $(v, \theta, \phi)$ in velocity space, and the polar axis $z$ is parallel to the field $\vec{B}_\mathrm{ext}$.

When a weak electric field $\vec{E}$ is applied to the plasma, we assume that the perturbation of the electron distribution function is also weak, $|f_1| \ll f_0$. We have to find the stationary current density associated with $f_1(t,\vec{r},\vec{v})$,
\begin{equation}
	\vec{j} = |q| n \hat\mu \vec{E} ,
\label{eq:Ohm}
\end{equation}
where $-\hat\mu$ is the local electron mobility tensor (with a minus sign) defined by nonlocal electron scattering on magnetic turbulence. 
In this linear approximation, Eq.~(\ref{eq:kin}) reduces to
\begin{equation}
	\frac{\partial f}{\partial t} + \vec{v} \nabla f - \frac{e f_0'}{m v}\vec{v}\vec{E} + \Omega\frac{\partial f_1}{\partial \phi} = \St[f_1],
\label{eq:kinLin}
\end{equation}
where $\Omega = eB_\mathrm{ext}/(mc)$ is the electron gyrofrequency in the external field, the derivative $f_0' = \partial f_0 / \partial v$, and for now we kept the space and time derivatives of $f$, neglecting only the term $\vec{E} \cdot \nabla_{\vec{v}} f_1$.
Bearing in mind the Einstein relations~\citep{Kubo1957} and averaging Eq.~(\ref{eq:kinLin}) over angles in velocity space under the conditions of Eq.~(\ref{eq:StProp}), we get
\begin{equation}
	\frac{\partial f_0}{\partial t} = - \nabla \int \! \frac{\vec{v}f_1}{4\pi} \, d\phi \, d\beta .
\label{eq:kinLinAver}
\end{equation}
Then, subtracting Eq.~(\ref{eq:kinLinAver}) from Eq.~(\ref{eq:kinLin}), we arrive at the equation for the distribution-function perturbation $f_1$ with an external force and a relaxation term:
\begin{equation}
    \frac{\partial f_1}{\partial t} + \vec{v} \nabla f_1 - \int \! \frac{\vec{v} \nabla f_1}{4\pi} \, d\phi \, d\beta + \vec{v} \left( \nabla f_0  - \frac{e f'_0}{m v} \vec{E} \right) = \left[ \St - \Omega \partial_\phi \right] f_1 \,.
\label{eq:kinLinf1}
\end{equation}

In the diffusion approximation (see, e.g.,~\citep{Webb2001, Subedi2017, Hasselmann1970}) the characteristic time and spatial scales of the perturbation of the distribution function, $T_\mathrm{c}$ and $L_\mathrm{c}$ respectively, are set to be larger than those of the operator on the right-hand side. In other words, when the inequalities $T_\mathrm{c}, \, v^{-1}L_\mathrm{c} \ \gg \ \Omega^{-1}, \, \tsc$ hold, where $\tsc$ is the typical scattering time, the first three terms on the left-hand side of Eq.~(\ref{eq:kinLinf1}) are negligible compared to the right-hand side:
\begin{equation}
    \left[ \St - \Omega \partial_\phi \right] f_1 = \vec{v} \left( \nabla f_0  - \frac{e f'_0}{m v} \vec{E} \right) \equiv \vec{v} \vec{\Pi}(t,\vec{r},v) .
\label{eq:kinLinDiffus}
\end{equation}

Thus, the plasma response, given by $f_1$, to the disturbance $\vec{\Pi}(t,\vec{r},v)$ is linear and dominated by turbulent scattering. Assuming the operator $\left[ \St - \Omega \partial_\phi \right]$ is invertible, we can express the function $f_1$ from Eq.~(\ref{eq:kinLinDiffus}) and insert it into Eq.~(\ref{eq:kinLinAver}) to obtain
\begin{equation}
	\frac{\partial f_0}{\partial t} -  \nabla\! \left(\hat{\kappa} \vec{\Pi}\right) = 0 .
\label{eq:contin}
\end{equation}
Here, in the absence of the electric field, $\vec{E} = 0$, we recognize the spatial diffusion equation for the velocity-isotropic component of the distribution function $f_0$ with the diffusion tensor
\begin{equation}
    \hat{\kappa} = \frac{1}{4\pi} \int\! \vec{v} \, \left[ \St - \Omega \partial_\phi \right]^{-1} \vec{v} \, d\phi d\beta .
\label{eq:diffusCoeff}
\end{equation}
When instead the spatial gradient is absent, $\nabla f_0 = 0$,  we multiply Eq.~(\ref{eq:contin}) by the particle charge and integrate over velocity space to arrive at what should be the charge continuity equation: $q \, \partial n / \partial t = - \nabla \vec{j}$. From Eq.~(\ref{eq:Ohm}) the integral form of the Einstein relation between the mobility and diffusion tensors (see, e.g., \cite{Kubo1957}) immediately follows:
\begin{equation}
	-\hat\mu = \frac{4\pi e}{m n} \int_0^{\infty} \! \hat\kappa f'_0 v dv .
\label{eq:Einstein}
\end{equation}

In the Cartesian coordinate system $(x, y, z)$, when the external magnetic field is parallel to the axis $z$, the mobility tensor has a form similar to that of the diffusion tensor:
\begin{equation}
    \hat\mu =
    \begin{pmatrix}
    \mu_\perp       & \mu_\mathrm {H} & 0 \\
    -\mu_\mathrm{H} & \mu_\perp      & 0 \\
    0                  & 0                 & \mu_\parallel
    \end{pmatrix} .
\label{eq:muMatrix}
\end{equation}
We compute its elements via a numerical procedure described below, using Eq.~(\ref{eq:Einstein}) for a Maxwellian equilibrium distribution of conducting electrons at a temperature $T$:
\begin{equation}
	f_\mathrm{M}(v) = n \left(\frac{m}{2\pi T}\right)^{3/2} \exp\left(-\frac{mv^2}{2T}\right).
    \label{eq:maxwell}
\end{equation}

By means of the Einstein relations Eq.~(\ref{eq:Einstein}), when valid, the three components of the electric conductivity tensor can be found from the three components of the diffusion tensor without a direct solution to the Boltzmann equation in velocity space, which may be more difficult compared to solving the particle diffusion problem in coordinate space. This approach allows one to avoid unnecessary simplifications like a so-called $\tau$-approximation of the collision (scattering) term, $ \St[f] = (f-f_0)\nu_\mathrm{s}$. 
At the same time, the famous Drude expression, $\sigma = -en\mu = (n e^2 /m)/\nu_s$, which is usually derived within the $\tau$-approximation, can be used for the interpretation of anisotropic anomalous conductivity and their comparison with the collisional ones on the basis of the effective scattering and collisional frequencies. Namely, we have simply $\mu = -(e/m)\tau_s$ for the longitudinal, transverse, and Hall electron mobility with the respective value of the mean free time $\tau_s = 1/\nu_s $.

\section{Evaluation of the Electron Diffusion and Mobility Tensors}

The components of the diffusion tensor for a fixed value of particle velocity $v$, in turn, can be obtained directly from simulating the test particles' motion through magnetostatic turbulence without any electric field applied. For the parallel and perpendicular components, $\kappa_\parallel$ and $\kappa_\perp$, we evaluate the diffusion coefficients from the mean-square displacement (MSD) of electrons propagating through the turbulence over sufficiently large time intervals $\Delta t$:
\begin{equation}
 \kappa_\parallel = \frac{\left\langle \Delta z^2 \right\rangle}{2 \Delta t} \, ,
 \quad
 \kappa_\perp = \frac{\left\langle \Delta x^2 + \Delta y^2 \right\rangle}{4 \Delta t} \,,
\end{equation}
where the angular brackets denote averaging over electron trajectories, and the time interval $\Delta t$ is assumed to be much greater than the scattering time $\tau_\mathrm{s}$, ensuring that the diffusive regime is attained.

In addition to the MSD approach, we verify the calculated diffusion coefficients $\kappa_\parallel$ and $\kappa_\perp$ by using the Taylor--Green--Kubo (TGK)~\citep{Kubo1957, Subedi2017, Shalchi2020, Kuhlen2025a} formalism, assuming ergodicity and stationarity of the turbulence. Within this framework, all diffusion coefficients are expressed in terms of the velocity autocorrelation function:
\begin{gather}
    \kappa_\parallel = \! \int_0^{\infty} \! \left\langle v_z(t) v_z(t+\tau) \right\rangle d\tau ,
    \label{eq:autocorrPar}
    \\
    \kappa_\perp = \frac{1}{2} \int_0^{\infty} \! \left\langle \vec{v}_\perp(t) \vec{v}_\perp(t+\tau) \right\rangle d\tau ,
    \label{eq:autocorrPerp}
    \\
    \kappa_\mathrm{H} = \frac{1}{2} \int_0^{\infty} \! \left\langle v_x(t)\,v_y(t+\tau) - v_y(t)\,v_x(t+\tau) \right\rangle d\tau.
    \label{eq:autocorrH}
\end{gather}
For a given time lag $\tau$, the autocorrelation function is evaluated by sampling multiple time origins $t$ along each particle trajectory with a fixed value of particle velocity $v$, computing the correlation between velocity components at times $t$ and $t+\tau$, and averaging over these samples. The resulting quantity is then averaged over an ensemble of $N$ electrons with random initial positions and velocity directions. Finally, the diffusion coefficients are obtained by numerical integration over $\tau$ as per Eqs.~(\ref{eq:autocorrPar})--(\ref{eq:autocorrH}). 

While both MSD and TGK approaches are used for the coefficients $\kappa_\parallel$ and $\kappa_\perp$, the antisymmetric (Hall) component $\kappa_\mathrm{H}$ of the diffusion tensor cannot be evaluated via MSD in a uniform external field, when particle drifts are absent, and therefore is determined here exclusively in the TGK formalism.

In the simulations, we use the Boris algorithm~\citep{Boris1970} to compute the trajectories of $N = 19200$ electrons propagating in the static magnetic field $\vec{B} = \vec{B}_\mathrm{ext} + \vec{B}_\mathrm{turb}$, defined on a grid, for a duration of up to $10^3 \, 2 \pi {\tilde\Omega}^{-1}$, where $\tilde\Omega$ is the electron gyrofrequency defined by the total mean square magnetic field. 
A set of $16$ initial velocities distributed evenly on a log scale over the interval from $v_\mathrm{min}=0.5 \, \Omega r_\mathrm{min}$ to $v_\mathrm{max}=2 \, \Omega r_\mathrm{max}$ is used, where $r_\mathrm{min}$ and $r_\mathrm{max}$ are the grid step and the simulation box sizes, respectively. (The box has periodic boundary conditions for particles.) 
In other words, the calculations have been carried out for a wide range of electron velocities from $v_\mathrm{min}$ to $v_\mathrm{max}$ in order to have a representative range ($3$ orders of magnitude) of the electron longitudinal and transversal rigidities,
\begin{equation}
\rho_{\parallel}=2\pi r_\mathrm{L}/L_{\mathrm{cor}\parallel},
\quad
\rho_{\perp}=2\pi r_\mathrm{L}/L_{\mathrm{cor}\perp},
\label{eq:rhos}
\end{equation}
respectively. Here $r_\mathrm{L}$ is the Larmor radius in the total mean square magnetic field, and the longitudinal and perpendicular correlation lengths, $L_{\mathrm{cor}\parallel}$ and $L_{\mathrm{cor}\perp}$, are defined below.

Spatial distributions of the magnetic field $\vec{B}$ are obtained from separate simulations carried out with the three-dimensional (3D) particle-in-cell ({PIC}) code {EPOCH}~\cite{Arber2015}. A {PIC} simulation is initiated in a simulation box with periodic boundary conditions, divided by $150 \times 150 \times 200$ cells, each of a size $r_\mathrm{min} = r_\mathrm{D}$, so that the maximum size $r_\mathrm{max} = 200 \, r_\mathrm{D}$. 
Here $r_\mathrm{D} = ( 2 T_\parallel/m )^{1/2} \omega_\mathrm{p}^{-1}$ is the Debye length, $\omega_\mathrm{p}$ is the plasma frequency.
The uniform external magnetic field $\vec{B}_\mathrm{ext}$ is parallel to the $z$ axis (and the largest side of the box), and the corresponding gyrofrequency is $\Omega/\omega_\mathrm{p} \times 10^2 = 0$, $2$, $3$, $6$, $9$, or $14$ in different runs. Electrons, represented by $800$ particles per cell, initially have a spatially homogeneous bi-Maxwellian velocity distribution with the anisotropy parameter $A = T_\parallel/T_\perp - 1 = 10$, where $T_\parallel$ and $T_\perp = 5.4 \times 10^{-3} \, m c^2$ are the temperatures parallel and orthogonal to the $z$ axis, respectively. Ions are motionless. Interparticle collisions are absent.

The Weibel-type instability~\cite{Weibel1959, Davidson1972, Emelyanov2024} of this anisotropic plasma leads to the exponential (aperiodic) growth of magnetic fields over a wide range of spatial scales, from several $r_\mathrm{min}$ to almost $r_\mathrm{max}$. The spatial spectrum of this turbulence is somewhat anisotropic and bears the same axial symmetry as the initial bi-Maxwellian electron distribution. After the instability saturates at a moment of time $t_\mathrm{sat}$, this magnetic turbulence evolves quasilinearly (see, e.g., \cite{Davidson1972, Kuznetsov2023}), gradually becomes more and more isotropic and slowly decays on a timescale $\tau_\mathrm{W} \gg r_\mathrm{max} / v_\mathrm{min}$.
This inequality justifies using the snapshots of magnetic field distributions $\vec{B} = \vec{B}_\mathrm{ext} + \vec{B}_\mathrm{turb}$, taken from {PIC} simulations at different time moments from $t_\mathrm{sat}$ to $10 \, t_\mathrm{sat}$, as static turbulence. Note that the external field impacts the spatial spectrum of the Weibel instability (see, e.g.,~\cite{Emelyanov2024, Pokhotelov2012, Ibscher2012}), so that $\vec{B}_\mathrm{turb} = \vec{B}_\mathrm{turb}(\vec{B}_\mathrm{ext})$, and hence the collision operator $\St$ in Eq.~(\ref{eq:diffusCoeff}) defining the diffusion tensor.

To find the correlation length of the turbulence, we first introduce the autocorrelation function of the turbulent magnetic field: $C(\vec{r}) = \left\langle B_\mathrm{turb}^2 \right\rangle^{-1} \left\langle \vec{B}_\mathrm{turb}(\vec{r}+\vec{x})\vec{B}_\mathrm{turb}(\vec{x}) \right\rangle$, where the angular brackets now denote integration over space, $\langle ... \rangle = \int\! ... \,d^3\vec{x}$. Then the correlation length along the axis $i = x,y,z$ is $L_{\mathrm{cor},i} = \int\! C(r_{j \neq i} = 0, r_i) \,dr_i$, and in the following, we use the quantities $L_{\mathrm{cor}\parallel} = L_{\mathrm{cor},z}$ and $L_{\mathrm{cor}\perp} = \left( L_{\mathrm{cor},x} + L_{\mathrm{cor},y} \right) / 2$. 

The ratios of these correlation lengths, as well as the average parallel and perpendicular wavenumbers and the average parallel and perpendicular spectrum widths, all normalized to the respective correlation lengths of the magnetic field, are summarized in Table~\ref{tbl:params}. They are shown for 10 typical turbulence snapshots characterized by different levels of magnetic fluctuations, which are defined by the ratio of the mean square of the turbulent magnetic field to the sum of this mean square value and the square value of the external magnetic field: 
\begin{equation}
\eta = \langle \vec{B}_\mathrm{turb}^2\rangle/(\vec{B}_\mathrm{ext}^2 + \langle \vec{B}_\mathrm{turb}^2\rangle).
\end{equation}

In the 1st and 9th rows of Table~\ref{tbl:params}, the external magnetic field is approximately 5 times larger and 5 times smaller, respectively, than the root-mean-square value of the turbulent field. For relatively strong external fields, i.e., for the upper half of Table~\ref{tbl:params}, the inverse values of the longitudinal and perpendicular correlation lengths are an order of magnitude smaller than the respective spectral widths. For relatively weak external fields, i.e., for the lower half of Table~\ref{tbl:params}, these longitudinal and perpendicular values, respectively, are more or less
close within a factor of order 2. For all values of the external field, the longitudinal average wavenumbers and spectral widths are close to each other (except for the last row with zero external field), but the perpendicular average wavenumbers and spectral widths differ by approximately two times (except for the first row with the strongest external field). Note that the average wavenumbers are situated on the long wavelength side of the spectrum and are about two to several times smaller than the short wavelength spectral boundary. As a rule, such a difference becomes greater for higher values of the external magnetic field in the cases of Table~\ref{tbl:params}. For times sufficiently longer than the saturation time $t_\mathrm{sat}$, in the considered case of a strong initial anisotropy $A = 10$ of the bi-Maxwellian distribution, a higher external field means a weaker anisotropy of the Weibel-type turbulence; i.e., the ratio $L_{\mathrm{cor}\parallel} / L_{\mathrm{cor}\perp}$ is closer to unity.

\section{Results and Discussion}

\begin{figure}[htbp!]
\includegraphics[width=\textwidth]{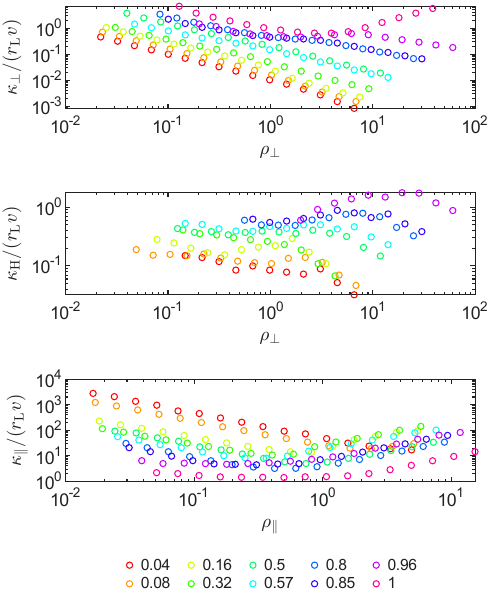}
\caption{Dependencies of the components of the normalized diffusion tensor $\hat\kappa$ on rigidity $\rho$ for different turbulence levels $\eta$ (see the legend). The different correlation lengths are used for rigidity normalization: $\rho_\perp=2\pi r_\mathrm{L}/L_{\mathrm{cor}\perp}$ and $\rho_\parallel=2\pi r_\mathrm{L}/L_{\mathrm{cor}\parallel}$}
\label{fig:kappaRig}
\end{figure}

The results of the numerical computation of three components of the diffusion tensor as functions of the respective perpendicular or longitudinal rigidities, $\rho_{\perp}$ or
$\rho_{\parallel}$ [see~\eqref{eq:rhos}], are presented in Fig.~\ref{fig:kappaRig} for different turbulence levels $\eta = \langle \vec{B}_\mathrm{turb}^2\rangle/(\vec{B}_\mathrm{ext}^2 + \langle \vec{B}_\mathrm{turb}^2\rangle)$. It is clear from the lower panel of the figure that for all values of the $\eta$ parameter, the dimensionless longitudinal coefficient of diffusion, normalized to the Bohm-type one~\citep{Casse2001}, i.e., the product of the electron Larmor radius, $r_{L}$, (calculated for the full mean square magnetic field) and the particle velocity, $v$, exhibits non-monotonic behavior with a prominent single minimum near the unit value of the rigidity. The visible variation of the minimum position in different curves reflects specific features of the turbulent spectrum, which depend on the time of its snapshot and the value of the external field. This variation can also be attributed to the fact that, for the computation of some of them, instantaneous snapshots of the turbulent field corresponding to the same Weibel instability simulation at different times after the saturation stage have been used. According to Table~\ref{tbl:dims}, the scaling law for the longitudinal diffusion coefficient in the limit of small rigidities, $\rho_{\parallel}\ll1$, can be approximated by a negative power law $\kappa_{\parallel}\propto\rho_{\parallel}^{\alpha_{\parallel1}}$, while for larger rigidities, $\rho_{\parallel}\gg1$, we have a positive power law $\kappa_{\parallel}\propto\rho_{\parallel}^{\alpha_{\parallel2}}$. Such a dependency is supposed to be a result of the corresponding different scaling laws of the effective collisional (scattering) frequency in these two limits (see, e.g., \citep{Bian2016, Reichherzer2025}). 

The general behavior of the perpendicular diffusion coefficient depends on the value of the relative turbulent level $\eta$ and, hence, on the external magnetic field $B_{ext}$ applied. For the zero external field, the longitudinal and perpendicular components coincide almost identically, representing a V-type shape. However, for the considerably strong field, i.e., for $\eta\lesssim0.7$, the perpendicular diffusion coefficient decreases monotonically as a result of the strong magnetization of electrons. The magnetization effect is also important for the Hall diffusion coefficient, which is shown in the intermediate panel of Fig.~\ref{fig:kappaRig} and demonstrates non-monotonic behavior at the perpendicular rigidities in the range of order 0.1 -- 10. For weaker external fields, this behavior is more prominent and shifts to higher values of rigidities.

It should be mentioned that the general shape of some obtained curves and the typical values of the diffusion coefficient in our nonrelativistic simulations resemble the results presented in the article~\citep{Casse2001} for the diffusion of relativistic particles in a turbulent field with a different set of properties. Nevertheless, in our simulations carried out for a variety of ten samples of magnetostatic turbulence, the differences in the obtained curves are quite prominent and cannot be reduced to those found in~\citep{Casse2001}.  

\begin{figure}[htbp!]
\includegraphics[width=\textwidth]{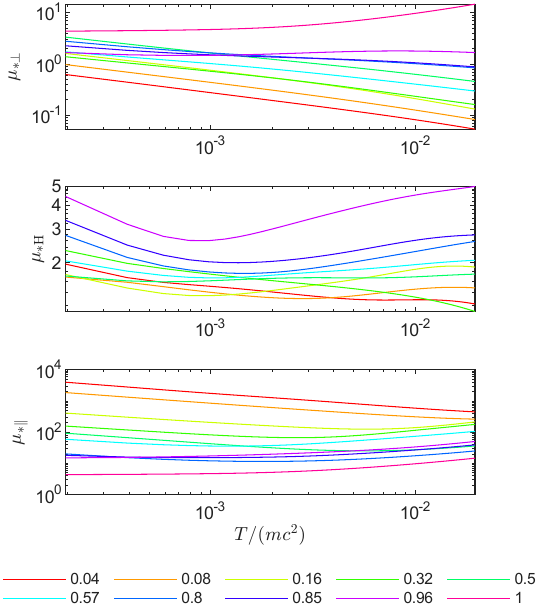}
\caption{Dependencies of the components of the normalized mobility tensor $\hat\mu_*$ on the normalized temperature $T/(mc^2)$ for different turbulence levels $\eta$ (see the legend).}
\label{fig:muTem}
\end{figure}

\begin{figure}[htbp!]
\includegraphics[width=\textwidth]{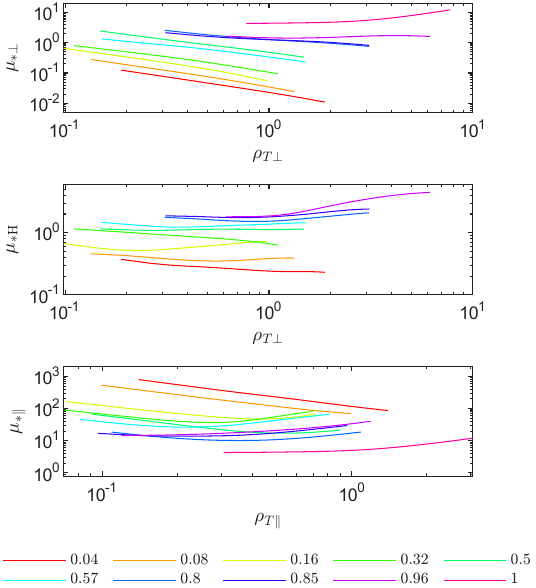}
\caption{Dependencies of the components of the normalized mobility tensor $\hat\mu_*$ on the thermal rigidity $\rho_T$ for different turbulence levels $\eta$ (see the legend). The different $\rho_T$ values are used: $\rho_{T\parallel}=2\pi v_T/\tilde\Omega L_{\mathrm{cor}\parallel}$ and $\rho_{T\perp}=2\pi v_T/\tilde\Omega L_{\mathrm{cor}\perp}$}
\label{fig:casseCmp}
\end{figure}

Using Einstein's relation~\eqref{eq:Einstein} and equations~\eqref{eq:muMatrix}, \eqref{eq:maxwell} in Sec.~\ref{sec:tensors}, we average the velocity-dependent diffusion coefficients over the derivative of the isotropic Maxwillian distribution and find the components of the mobility tensor $\hat{\mu}_*$, presented in Fig.~\ref{fig:muTem}-\ref{fig:casseCmp} in a normalized form according to the equality
\begin{equation}
\hat{\mu}_* (\rho_T) = \frac{m \tilde\Omega}{e}\hat\mu = \frac{8 }{\pi^{1/2}} \int_0^{\infty} \! \frac{\hat\kappa(\rho)}{r_\mathrm{L} v}  \exp\left(-\frac{\rho^2}{\rho_T^2}\right) \frac{\rho^4 d\rho}{\rho_T^5} \,,
\quad \rho_T = \frac{2\pi v_T}{\tilde\Omega L_\mathrm{cor}}.
\end{equation}
Here, the thermal rigidity is introduced for the thermal velocity of the Maxwellian distribution.

Figure~\ref{fig:muTem} shows the temperature dependencies of the mobility tensor components in the range $0.1 \div 10$~keV for different turbulence levels $\rho$. The general behavior of the curves, as expected, mirrors the corresponding ones for the diffusion tensor components in Fig.~\ref{fig:kappaRig}. For example, the perpendicular mobility decreases monotonically at almost any $\rho$, except for its value very close to unity when the external magnetic field is much weaker than the mean magnitude of the turbulent field. In this limit, the perpendicular mobility reaches its maximum. The values of this mobility at 0.1 and 10~keV can differ by several times (when the turbulence level $\eta$ is close to 1) to an order of magnitude (when $\eta$ is small). The longitudinal mobility has almost an opposite temperature dependence and its values at 0.1 and 10~keV can differ even more strongly, by two orders of magnitude. The Hall mobility shows non-monotonic temperature dependence, similar to that of the Hall diffusion coefficient, and the values of this mobility at 0.1 and 10~keV differ utmost in a couple of times.

Figure~\ref{fig:casseCmp} shows the three normalized mobility components as the functions of the thermal rigidity. Again, they all remind the respective diffusion coefficients as the function of rigidity in Fig.~\ref{fig:kappaRig} for the same turbulence levels $\eta$, but have smoother profiles due to averaging over thermal velocity distribution and a bit larger spreading along the thermal rigidity axis.

\begin{figure}[htbp!]
\includegraphics[width=\textwidth]{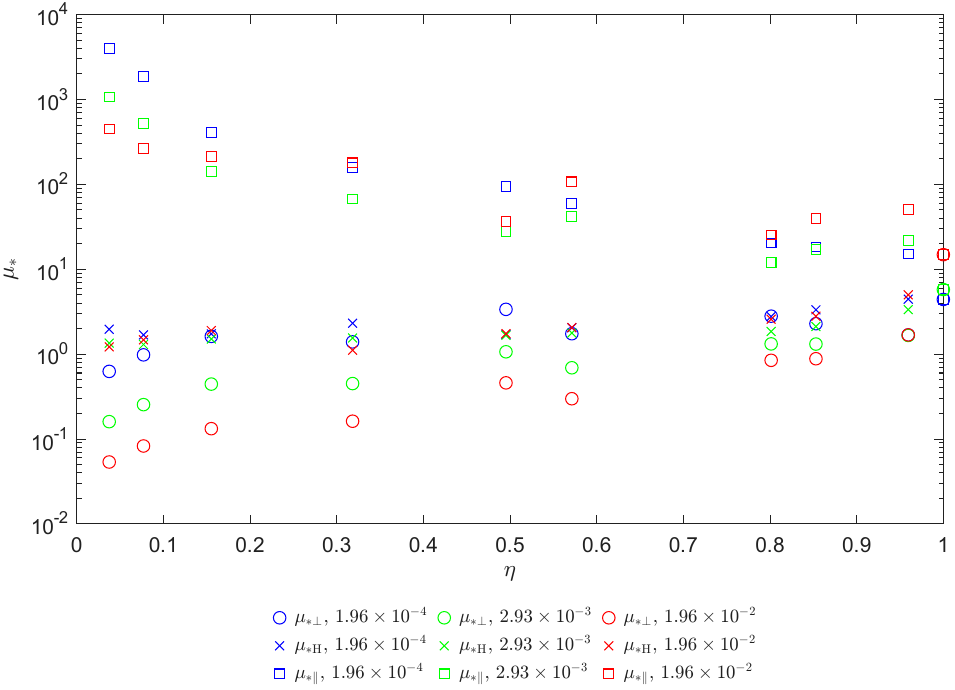}
\caption{Dependencies of the components of the normalized mobility tensor $\hat\mu_*$ on turbulence level $\eta$ for three values of the normalized temperature $T/(mc^2)$ (see the legend).}
\label{fig:muEta}
\end{figure}

The dependence of the perpendicular, longitudinal, and Hall mobilities on the $\rho$ parameter is stronger at higher, lower, and intermediate temperatures, respectively. Fig.~\ref{fig:muEta} shows the dependencies of the normalized mobility components $\hat{\mu}_*$ on the turbulence level $\eta$ for three values of electron temperature, 0.1, 1.0, and 10~keV. Most of these dependencies are slightly non-monotonic, according to the discussion above. Overall, the dependence on the turbulence level is quite strong: the values of the $\hat{\mu}_*$ tensor components at 0.1 and 10~keV can differ by more than two orders of magnitude.

We also calculated the temperature dependencies of the diagonal components of the normalized mobility tensor $\hat{\mu}_*$ for the data from~\citep{Casse2001} and compared them with ours for different turbulence levels $\eta$. The general patterns of curve behavior as a function of $\rho_T$ are the same in both cases. This confirms the validity of our results, which nevertheless demonstrate a larger variety of patterns due to the rich set of turbulence samples.

Note that the dimensionless longitudinal mobility tensor component introduced above equals the dimensionless effective scattering time, or inverse collision frequency $\nu_\mathrm{s}$, i.e., $\mu_{*\parallel}\equiv\ \tilde\Omega/\nu_{s\parallel}$. For our data, we have $2\pi\nu_\mathrm{s\parallel}/\tilde\Omega \sim 10^{-2}\div 10^2$, so the mean free time can be either less or greater than a gyroperiod. Also, we can estimate the electron mean free path as $\lambda = v_T/\nu_\mathrm{s\parallel}$; say, for $T=1$~keV it is equal $\lambda \sim (10^{-1} \div 10^{2}) L_{\mathrm{cor}\parallel}$ and can be either less or greater than the longitudinal correlation length.
% The dependency of the effective scattering rate on the electron rigidity can be obtained from the given by a simple inversion of it.

Although energy range of electrons in our simulations is more typical for extreme coronal plasma conditions, which usually occur in big solar flares, the obtained results are of great interest for the interpretation of physical processes in corona. Among others, an example is the problem of anomalous ohmic heating by return electric current in the presence of magnetostatic turbulence, which has been widely discussed, e.g., in \citep{Bian2016,emslieEffectsTurbulentElectrical2024, emslieEffectsTurbulentElectrical2025}, without any specification of the nature of the magnetic turbulence. In this respect, it can also be noted, that for more typical coronal parameters, i.e., plasma  temperature $T\sim \text{100 – 1000 ~eV}$, plasma density $n\sim 10^{8}-10^{10} \text{cm}^{-3}$, and average magnetic field $B_{ext}\sim 10 G$, in the case of Weibel-type magnetic turbulence the expected rigidity of thermal particles is about 0.1 – 1, and the corresponding anomalous collision frequency is $10^6-10^8\text{c}^{-1}$. Thereby, the dominance of the turbulent scattering over the Coulomb collisions can vary from three to five orders of magnitude, or even more for higher temperatures.

The mobility tensor components computed in this work can provide principle information for further research on the problem of anomalous plasma conductivity associated with quasi-magnetostatic kinetic turbulence and its manifestations in different astrophysical environments. They can be used to formulate the quasi-analytical models of the process or be incorporated into MHD simulations, which are quite useful since full kinetic or hybrid simulations of large magneto-plasma structures, like coronal loops or certain parts of them, are still complicated, computationally expensive, or impossible altogether. 

\section{Final Comments and Conclusion}

In this work we investigated anisotropic diffusion of nonrelativistic electrons and their mobility in collisionless plasma with homogeneous external magnetic field and quasi-static magnetic microturbulence, generated via kinetic Weibel-type instability. We obtained analytical expressions for the mobility tensor, determined by electron interaction with small-scale inhomogeneous magnetic field, and provided numerical computation of this tensor using an original code for finding the electron diffusion coefficients, based on the well-known Boris algorithm. The dependencies of the mobility tensor components on the electron temperature, magnetic turbulence level, external magnetic field, and electron rigidity were found.

The qualitative results obtained in this work can be applied to cosmic plasmas with a broad range of parameters, including stellar atmospheres and winds, accretion disks and jets, planetary magnetospheres and other objects, where quasi-magnetostatic microturbulence commonly develops. Nevertheless, some important questions, which are significant to the applicability of these results to actual processes, are remained out of the scope of this article. In particular, an additional analysis is required for: i) the validity of the quasi-stationary regime for magnetic field fluctuations in an actual environment, ii) the support of the turbulent field at a quasi-stationary level, iii) the interconnection of magnetic perturbations with other types of turbulence, iv) the backward influence of scattered particles on the turbulent magnetic field and its spectrum, v) the role of collective effects in the process of particle transport, vi) the effects of turbulence anisotropy, etc.

As is shown in the article, magnetic microturbulence can enhance plasma resistivity by many orders of magnitude. So, its role in the anomalous ohmic heating and dissipation of electric current structures in cosmic plasmas and the development of various astrophysical phenomena, related to reconnection processes, can be more significant than it was considered previously. The development of the fast magnetic reconnection models with a self-consistent kinetic quasi-magnetostatic turbulence is of particular interest for the interpretation of various processes in the solar corona, full of electric current structures of a broad range of scales and plasma parameters. Such a rec onnection can release the stored magnetic energy through explosive events of a large energy variety: from nano- and pico-flares up to gigantic bursts. Most of the open questions mentioned above will be the subject for future research, which are needed in further support from the numerical computations, especially multi-scale, experimental data, and space observations of
various kinetic effects related to the anomalous conductivity of a magnetoactive plasma.

\begin{acknowledgments}
This work was supported by the ``BASIS'' Foundation (project No.~24-1-1-97-1).
\end{acknowledgments}

\bibliography{bibliography.bib}

@article{Kontar2011,
  title = {Acceleration, Magnetic Fluctuations, and Cross-Field Transport of Energetic Electrons in a Solar Flare Loop},
  volume = {730},
  ISSN = {2041-8213},
  url = {http://dx.doi.org/10.1088/2041-8205/730/2/L22},
  doi = {10.1088/2041-8205/730/2/l22},
  number = {2},
  journal = {The Astrophysical Journal},
  publisher = {American Astronomical Society},
  author = {Kontar,  E. P. and Hannah,  I. G. and Bian,  N. H.},
  year = {2011},
  month = Mar,
  pages = {L22}
}

@book{Shalchi2009,
  title = {Nonlinear Cosmic Ray Diffusion Theories},
  url = {http://dx.doi.org/10.1007/978-3-642-00309-7},
  publisher = {Springer},
  author = {Shalchi,  Andreas},
  year = {2009}
}

@article{Sen2025,
  title = {Role of magnetic shear distribution in the formation of eruptive flux ropes},
  volume = {703},
  ISSN = {1432-0746},
  url = {http://dx.doi.org/10.1051/0004-6361/202556232},
  doi = {10.1051/0004-6361/202556232},
  journal = {Astronomy \& Astrophysics},
  publisher = {EDP Sciences},
  author = {Sen,  Samrat and Nayak,  Sushree S. and Antolin,  Patrick},
  year = {2025},
  month = Nov,
  pages = {A241}
}

@article{Dahlin2022,
  title = {Variability of the Reconnection Guide Field in Solar Flares},
  volume = {932},
  ISSN = {1538-4357},
  url = {http://dx.doi.org/10.3847/1538-4357/ac6e3d},
  doi = {10.3847/1538-4357/ac6e3d},
  number = {2},
  journal = {The Astrophysical Journal},
  publisher = {American Astronomical Society},
  author = {Dahlin,  Joel T. and Antiochos,  Spiro K. and Qiu,  Jiong and DeVore,  C. Richard},
  year = {2022},
  month = jun,
  pages = {94}
}

@article{Arnold2021,
  title = {Electron Acceleration during Macroscale Magnetic Reconnection},
  volume = {126},
  ISSN = {1079-7114},
  url = {http://dx.doi.org/10.1103/PhysRevLett.126.135101},
  doi = {10.1103/physrevlett.126.135101},
  pages = {13},
  journal = {Physical Review Letters},
  publisher = {American Physical Society (APS)},
  author = {Arnold,  H. and Drake,  J. F. and Swisdak,  M. and Guo,  F. and Dahlin,  J. T. and Chen,  B. and Fleishman,  G. and Glesener,  L. and Kontar,  E. and Phan,  T. and Shen,  C.},
  year = {2021},
  month = Mar 
}

@article{Aschwanden2019,
  title = {Global Energetics of Solar Flares. VIII. The Low-energy Cutoff},
  volume = {881},
  ISSN = {1538-4357},
  url = {http://dx.doi.org/10.3847/1538-4357/ab2cd4},
  doi = {10.3847/1538-4357/ab2cd4},
  number = {1},
  journal = {The Astrophysical Journal},
  publisher = {American Astronomical Society},
  author = {Aschwanden,  Markus J. and Kontar,  Eduard P. and Jeffrey,  Natasha L. S.},
  year = {2019},
  month = Aug,
  pages = {1}
}

@article{Ji2007,
  title = {The Relaxation of Sheared Magnetic Fields: A Contracting Process},
  volume = {660},
  ISSN = {1538-4357},
  url = {http://dx.doi.org/10.1086/513017},
  doi = {10.1086/513017},
  number = {1},
  journal = {The Astrophysical Journal},
  publisher = {American Astronomical Society},
  author = {Ji,  Haisheng and Huang,  Guangli and Wang,  Haimin},
  year = {2007},
  month = May,
  pages = {893–900}
}

@article{Rosenbluth1966,
  title = {Destruction of magnetic surfaces by magnetic field irregularities},
  volume = {6},
  doi = {10.1088/0029-5515/6/4/008},
  number = {4},
  journal = {Nuclear Fusion},
  author = {Rosenbluth,  M. N. and Sagdeev,  R. Z. and Taylor,  J. B. and Zaslavski,  G. M.},
  year = {1966},
  month = dec,
  pages = {297--300}
}

@article{emslieEffectsTurbulentElectrical2024,
  title = {The {{Effects}} of {{Turbulent Electrical Resistivity}} on the {{Response}} of the {{Solar Atmosphere}} to {{Flare Energy Input}}. {{I}}. {{Results}} of {{Radiative Hydrodynamic Simulations}}},
  author = {Emslie, A. Gordon and Allred, Joel C. and Alaoui, Meriem},
  year = {2024},
  month = dec,
  journal = {The Astrophysical Journal},
  volume = {977},
  number = {2},
  pages = {246},
  issn = {0004-637X, 1538-4357},
  doi = {10.3847/1538-4357/ad919c},
  urldate = {2026-04-28},  
}

@article{emslieEffectsTurbulentElectrical2025,
  title = {The {{Effects}} of {{Turbulent Electrical Resistivity}} on the {{Response}} of the {{Solar Atmosphere}} to {{Flare Energy Input}}. {{II}}. {{Hard X-Ray Spectra}}, {{Light Curves}}, and {{Spatial Distributions}}},
  author = {Emslie, A. Gordon and Allred, Joel C. and Alaoui, Meriem},
  year = {2025},
  month = nov,
  journal = {The Astrophysical Journal},
  volume = {993},
  number = {1},
  pages = {127},
  issn = {0004-637X, 1538-4357},
  doi = {10.3847/1538-4357/ae0709},
  urldate = {2026-04-28}
}

@book{Biskamp1993,
  title     = "A Nonlinear Magnetohydrodynamics",
  author    = "Biskamp, D.",
  year      =  1993,
  publisher = "Cambridge University Press",
  address   = "Cambridge"
}

@article{Krommes2002,
  title = {Fundamental statistical descriptions of plasma turbulence in magnetic fields},
  volume = {360},
  ISSN = {0370-1573},
  url = {http://dx.doi.org/10.1016/S0370-1573(01)00066-7},
  doi = {10.1016/s0370-1573(01)00066-7},
  number = {1--4},
  journal = {Physics Reports},
  publisher = {Elsevier BV},
  author = {Krommes,  John A.},
  year = {2002},
  month = apr,
  pages = {1--352}
}

@article{Matthaeus2011,
  title = {Who Needs Turbulence?: {A} Review of Turbulence Effects in the Heliosphere and on the Fundamental Process of Reconnection},
  volume = {160},
  doi = {10.1007/s11214-011-9793-9},
  number = {1--4},
  journal = {Space Science Reviews},
  publisher = {Springer Science and Business Media LLC},
  author = {Matthaeus,  W. H. and Velli,  M.},
  year = {2011},
  month = jun,
  pages = {145--168}
}

@article{Karimabadi2013,
  title = {Magnetic reconnection in the presence of externally driven and self-generated turbulence},
  volume = {20},
  doi = {10.1063/1.4828395},
  pages = {11},
  journal = {Physics of Plasmas},
  author = {Karimabadi,  H. and Lazarian,  A.},
  year = {2013},
  month = nov 
}

@article{Matthaeus2021,
  title = {Turbulence in space plasmas: {W}ho needs it?},
  volume = {28},
  doi = {10.1063/5.0041540},
  number = {3},
  journal = {Physics of Plasmas},
  author = {Matthaeus,  W. H.},
  year = {2021},
  month = mar,
  pages = {032306}
}

@article{Vogman2025,
  title = {Parameterized anomalous transport model for current-carrying collisionless plasmas in pulsed power inertial confinement fusion},
  volume = {32},
  doi = {10.1063/5.0278612},
  pages={082305},
  number = {8},
  journal = {Physics of Plasmas},
  author = {Vogman,  G. V. and Ho,  A. and Hammer,  J. H.},
  year = {2025},
  month = aug 
}

@article{Graham2025,
  title = {The Role of Kinetic Instabilities and Waves in Collisionless Magnetic Reconnection},
  volume = {221},
  doi = {10.1007/s11214-024-01133-7},
  number = {1},
  pages={20},
  journal = {Space Science Reviews},
  author = {Graham,  D. B. and Cozzani,  G. and Khotyaintsev,  Yu. V. and Wilder,  V. D. and Holmes,  J. C. and Nakamura,  T. K. M. and B\"{u}chner,  J. and Dokgo,  K. and Richard,  L. and Steinvall,  K. and Norgren,  C. and Chen,  L.-J. and Ji,  H. and Drake,  J. F. and Stawarz,  J. E. and Eriksson,  S.},
  year = {2025},
  month = feb 
}

@article{Norman1978,
       author = {Norman, C.~A. and Smith, R.~A.},
        title = "Kinetic processes in solar flares",
      journal = {Astronomy \& Astrophysics},
         year = 1978,
        month = aug,
       volume = {68},
       number = {1--2},
        pages = {145--155}
}

@article{Bian2016,
  title = {Suppression of parallel transport in turbulent magnetized plasmas and its impact on the non-thermal and thermal aspects of solar flares},
  volume = {824},
  doi = {10.3847/0004-637x/824/2/78},
  number = {2},
  journal = {The Astrophysical Journal},
  author = {Bian,  Nicolas H. and Kontar,  Eduard P. and Emslie,  A. Gordon},
  year = {2016},
  month = jun,
  pages = {78}
}

@article{Yoon2017,
  title = {Kinetic instabilities in the solar wind driven by temperature anisotropies},
  volume = {1},
  doi = {10.1007/s41614-017-0006-1},
  pages={4},
  number = {1},
  journal = {Reviews of Modern Plasma Physics},
  publisher = {Springer Science and Business Media LLC},
  author = {Yoon,  Peter H.},
  year = {2017},
  month = jul
}

@article{Bian2018,
  title = {Heating and Cooling of Coronal Loops with Turbulent Suppression of Parallel Heat Conduction},
  volume = {852},
  doi = {10.3847/1538-4357/aa9f29},
  number = {2},
  journal = {The Astrophysical Journal},
  publisher = {American Astronomical Society},
  author = {Bian,  Nicolas and Emslie,  A. Gordon and Horne,  Duncan and Kontar,  Eduard P.},
  year = {2018},
  month = jan,
  pages = {127}
}

@article{Khotyaintsev2019,
  title = {Collisionless Magnetic Reconnection and Waves: Progress Review},
  author = {Khotyaintsev, Yuri V. and Graham, Daniel B. and Norgren, Cecilia and Vaivads, Andris},
  year = {2019},
  journal = {Frontiers in Astronomy and Space Sciences},
  volume = {6},
  pages = {70},
  doi = {10.3389/fspas.2019.00070},
}

@article{Yoon2019,
  title = {Solar Wind Temperature Isotropy},
  volume = {123},
  doi = {10.1103/physrevlett.123.145101},
  number = {14},
  pages={145101},
  journal = {Physical Review Letters},
  author = {Yoon,  P. H. and Seough,  J. and Salem,  C. S. and Klein,  K.G.},
  year = {2019},
  month = oct 
}

@article{Graham2022,
  title = {Direct observations of anomalous resistivity and diffusion in collisionless plasma},
  volume = {13},
  doi = {10.1038/s41467-022-30561-8},
  number = {1},
  pages={2954},
  journal = {Nature Communications},
  author = {Graham,  D. B. and Khotyaintsev,  Yu. V. and Andr\'{e},  M. and Vaivads,  A. and Divin,  A. and Drake,  J. F. and Norgren,  C. and Le Contel,  O. and Lindqvist,  P.-A. and Rager,  A. C. and Gershman,  D. J. and Russell,  C. T. and Burch,  J. L. and Hwang,  K.-J. and Dokgo,  K.},
  year = {2022},
  month = may 
}

@article{Isliker2025,
  title = {Transport of particles in strongly turbulent 3D magnetized plasmas},
  volume = {32},
  doi = {10.1063/5.0285560},
  pages={090501},
  number = {9},
  journal = {Physics of Plasmas},
  publisher = {AIP Publishing},
  author = {Isliker,  Heinz and Vlahos,  Loukas},
  year = {2025},
  month = sep 
}

@article{Rostoker1961,
  title = {Fluctuations of a plasma (I)},
  volume = {1},
  doi = {10.1088/0029-5515/1/2/004},
  number = {2},
  journal = {Nuclear Fusion},
  author = {Rostoker,  Norman},
  year = {1961},
  month = mar,
  pages = {101--120}
}

@book{Akhiezer1975,
  title     = "Plasma Electrodynamics. 2: {N}on-linear Theory and Fluctuations",
  author    = "Akhiezer, A. I.",
  year      =  1975,
  publisher = "Pergamon Press",
  address   = "Oxford"
}

@article{Che2017,
  title = {How Anomalous Resistivity Accelerates Magnetic Reconnection},
  author = {Che, H.},
  year = {2017},
  journal = {Physics of Plasmas},
  volume = {24},
  number = {8},
  pages = {082115},
  doi = {10.1063/1.5000071}
}

@article{Yoo2025,
  title = {Anomalous Resistivity and Electron Heating by Lower Hybrid Drift Waves inside Reconnecting Current Sheets},
  author = {Yoo, Jongsoo and Ji, Hantao and Shi, Peiyun and Bose, Sayak and Ng, Jonathan and Chen, Li-Jen and Yamada, Masaaki},
  year = {2025},
  journal = {Physics of Plasmas},
  volume = {32},
  number = {6},
  pages = {062114},
  doi = {10.1063/5.0271730}
}

@article{Carter2002,
  title = {Experimental study of lower-hybrid drift turbulence in a reconnecting current sheet},
  volume = {9},
  doi = {10.1063/1.1494433},
  number = {8},
  journal = {Physics of Plasmas},
  author = {Carter,  T. A. and Yamada,  M. and Ji,  H. and Kulsrud,  R. M. and Trintchouk,  F.},
  year = {2002},
  month = aug,
  pages = {3272--3288}
}

@article{Ji2004,
  title = {Electromagnetic Fluctuations during Fast Reconnection in a Laboratory Plasma},
  volume = {92},
  doi = {10.1103/physrevlett.92.115001},
  pages={115001},
  number = {11},
  journal = {Physical Review Letters},
  author = {Ji,  Hantao and Terry,  Stephen and Yamada,  Masaaki and Kulsrud,  Russell and Kuritsyn,  Aleksey and Ren,  Yang},
  year = {2004},
  month = mar 
}

@article{Ji2023,
  title = {Laboratory Study of Collisionless Magnetic Reconnection},
  author = {Ji, H. and Yoo, J. and Fox, W. and Yamada, M. and Argall, M. and Egedal, J. and Liu, Y.-H. and Wilder, R. and Eriksson, S. and Daughton, W. and Bergstedt, K. and Bose, S. and Burch, J. and Torbert, R. and Ng, J. and Chen, L.-J.},
  year = {2023},
  journal = {Space Science Reviews},
  volume = {219},
  number = {8},
  pages = {76},
  doi = {10.1007/s11214-023-01024-3}
}

@article{Daughton2011,
  title = {Role of electron physics in the development of turbulent magnetic reconnection in collisionless plasmas},
  volume = {7},
  doi = {10.1038/nphys1965},
  number = {7},
  journal = {Nature Physics},
  author = {Daughton,  W. and Roytershteyn,  V. and Karimabadi,  H. and Yin,  L. and Albright,  B. J. and Bergen,  B. and Bowers,  K. J.},
  year = {2011},
  month = apr,
  pages = {539--542}
}

@article{Tummel2020,
  title = {Kinetic simulations of anomalous resistivity in high-temperature current carrying plasmas},
  volume = {27},
  doi = {10.1063/5.0004508},
  number = {9},
  journal = {Physics of Plasmas},
  author = {Tummel,  K. and Ellison,  C. L. and Farmer,  W. A. and Hammer,  J. H. and Parker,  J. B. and LeChien,  K. R.},
  year = {2020},
  month = sep,
  pages = {092306} 
}

@article{Wang2021,
  title = {Lower-Hybrid Drift Waves and Their Interaction with Plasmas in a {{3D}} Symmetric Reconnection Simulation with Zero Guide Field},
  author = {Wang, Shan and Chen, Li-Jen and Ng, Jonathan and Bessho, Naoki and Hesse, Michael},
  year = {2021},
  journal = {Physics of Plasmas},
  volume = {28},
  number = {7},
  pages = {072102},
  doi = {10.1063/5.0054626},
}

@article{Nechaev2022,
  title = {Turbulent multicomponent magnetopause: Analytical description and kinetic simulation of distributed current sheets},
  volume = {32},
  doi = {10.1063/5.0254381},
  pages={082901},
  number = {8},
  journal = {Physics of Plasmas},
  author = "Nechaev,  A. A. and Garasev,  M. A. and Kocharovsky,  {\relax Vl}. V.",
  year = {2022},
  month = aug 
}

@article{Vicentin2025,
  title = {Investigating Turbulence Effects on Magnetic Reconnection Rates through 3D Resistive Magnetohydrodynamic Simulations},
  volume = {987},
  doi = {10.3847/1538-4357/addc62},
  number = {2},
  journal = {The Astrophysical Journal},
  author = {Vicentin,  Giovani H. and Kowal,  Grzegorz and de Gouveia Dal Pino,  Elisabete M. and Lazarian,  Alex},
  year = {2025},
  month = 7,
  pages = {213}
}

@article{Ergun2020,
  title = {Particle Acceleration in Strong Turbulence in the Earth's Magnetotail},
  volume = {898},
  doi = {10.3847/1538-4357/ab9ab5},
  number = {2},
  journal = {The Astrophysical Journal},
  author = {Ergun,  R. E. and Ahmadi,  N. and Kromyda,  L. and Schwartz,  S. J. and Chasapis,  A. and Hoilijoki,  S. and Wilder,  F. D. and Cassak,  P. A. and Stawarz,  J. E. and Goodrich,  K. A. and Turner,  D. L. and Pucci,  F. and Pouquet,  A. and Matthaeus,  W. H. and Drake,  J. F. and Hesse,  M. and Shay,  M. A. and Torbert,  R. B. and Burch,  J. L.},
  year = {2020},
  month = aug,
  pages = {153}
}

@article{Cozzani2021,
  title = {Structure of a {{Perturbed Magnetic Reconnection Electron Diffusion Region}} in the {{Earth}}'s {{Magnetotail}}},
  author = {Cozzani, G. and Khotyaintsev, Yu. V. and Graham, D. B. and Egedal, J. and Andr\'{e}, M. and Vaivads, A. and Alexandrova, A. and Le Contel, O. and Nakamura, R. and Fuselier, S. A. and Russell, C. T. and Burch, J. L.},
  year = {2021},
  journal = {Physical Review Letters},
  volume = {127},
  number = {21},
  pages = {215101},
  doi = {10.1103/PhysRevLett.127.215101}
}

@article{Lotekar2022,
  title = {Kinetic-Scale Current Sheets in Near-Sun Solar Wind: Properties, Scale-dependent Features and Reconnection Onset},
  author = {Lotekar, A. B. and Vasko, I. Y. and Phan, T. and Bale, S. D. and Bowen, T. A. and Halekas, J. and Artemyev, A. V. and Khotyaintsev, Yu. V. and Mozer, F. S.},
  year = 2022,
  journal = {The Astrophysical Journal},
  volume = {929},
  number = {1},
  pages = {58},
  doi = {10.3847/1538-4357/ac5bd9},
}

@article{Howes2024,
  title = {The Fundamental Parameters of Astrophysical Plasma Turbulence and Its Dissipation: Non-Relativistic Limit},
  author = {Howes, Gregory G.},
  year = {2024},
  journal = {Journal of Plasma Physics},
  volume = {90},
  number = {5},
  pages = {905900504},
  doi = {10.1017/S0022377824001090},
}

@article{Shibata2011,
  title = {Solar Flares: Magnetohydrodynamic Processes},
  volume = {8},
  pages={6},
  doi = {10.12942/lrsp-2011-6},
  journal = {Living Reviews in Solar Physics},
  author = {Shibata,  Kazunari and Magara,  Tetsuya},
  year = {2011}
}

@article{Benz2016,
  title = {Flare Observations},
  volume = {14},
  doi = {10.1007/s41116-016-0004-3},
  pages = {2},
  number = {1},
  journal = {Living Reviews in Solar Physics},
  publisher = {Springer Science and Business Media LLC},
  author = {Benz,  Arnold O.},
  year = {2016},
  month = dec 
}

@article{Gritsyk2021,
  title = {Modern analytic models of acceleration and propagation of electrons in solar flares},
  volume = {66},
  ISSN = {1468-4780},
  url = {http://dx.doi.org/10.3367/UFNe.2021.08.039048},
  doi = {10.3367/ufne.2021.08.039048},
  number = {05},
  journal = {Physics-Uspekhi},
  author = {Gritsyk,  Pavel A. and Somov,  Boris V.},
  year = {2021},
  month = aug,
  pages = {437--459}
}

@article{Parker1983a,
  title = {Magnetic neutral sheets in evolving fields. I~- General theory. II~- Formation of the solar corona},
  volume = {264},
  doi = {10.1086/160636},
  journal = {The Astrophysical Journal},
  author = {Parker,  E. N.},
  year = {1983},
  month = jan,
  pages = {635--647}
}

@article{Emslie2018,
  title = {Reduction of Thermal Conductive Flux by Non-local Effects in the Presence of Turbulent Scattering},
  author = {Emslie, A. Gordon and Bian, N. H.},
  year = {2018},
  journal = {The Astrophysical Journal},
  volume = {865},
  number = {1},
  pages = {67},
  doi = {10.3847/1538-4357/aad961}
}

@article{Lazarian2020,
  title = "{3D} Turbulent Reconnection: Theory, Tests, and Astrophysical Implications",
  author = {Lazarian, Alex and Eyink, Gregory L. and Jafari, Amir and Kowal, Grzegorz and Li, Hui and Xu, Siyao and Vishniac, Ethan T.},
  year = {2020},
  journal = {Physics of Plasmas},
  volume = {27},
  number = {1},
  pages = {012305},
  doi = {10.1063/1.5110603}
}

@article{Medvedev2000,
  title = {Theory of ``Jitter'' Radiation from Small-Scale Random Magnetic Fields and Prompt Emission from Gamma-Ray Burst Shocks},
  volume = {540},
  doi = {10.1086/309374},
  number = {2},
  journal = {The Astrophysical Journal},
  publisher = {American Astronomical Society},
  author = {Medvedev,  Mikhail V.},
  year = {2000},
  month = sep,
  pages = {704--714}
}

@article{Casse2001,
 title = {Transport of cosmic rays in chaotic magnetic fields},
 volume = {65},
 ISSN = {1089-4918},
 url = {http://doi.org/10.1103/PhysRevD.65.023002},
 doi = {10.1103/physrevd.65.023002},
 number = {2},
 journal = {Physical Review D},
 publisher = {American Physical Society (APS)},
 author = {Casse, Fabien and Lemoine, Martin and Pelletier, Guy},
 year = {2001},
 month = nov,
 pages={023002},
}

@article{Plotnikov2011,
  title = {Particle transport in intense small-scale magnetic turbulence with a mean field},
  volume = {532},
  doi = {10.1051/0004-6361/201117182},
  journal = {Astronomy \& Astrophysics},
  author = {Plotnikov,  I. and Pelletier,  G. and Lemoine,  M.},
  year = {2011},
  month = jul,
  pages = {A68}
}

@article{Keenan2013,
  title = {Particle transport and radiation production in sub-Larmor-scale electromagnetic turbulence},
  volume = {88},
  ISSN = {1550-2376},
  url = {http://dx.doi.org/10.1103/PhysRevE.88.013103},
  doi = {10.1103/physreve.88.013103},
  number = {1},
  pages={013103},
  journal = {Physical Review E},
  publisher = {American Physical Society (APS)},
  author = {Keenan,  Brett D. and Medvedev,  Mikhail V.},
  year = {2013},
  month = 7 
}

@article{Shalchi2020,
  title = {Perpendicular Transport of Energetic Particles in Magnetic Turbulence},
  volume = {216},
  doi = {10.1007/s11214-020-0644-4},
  number = {2},
  journal = {Space Science Reviews},
  publisher = {Springer Science and Business Media LLC},
  author = {Shalchi,  Andreas},
  year = {2020},
  month = feb,
  pages={23},
}

@article{Zhang2024,
 title = {Cosmic Ray Diffusion in Magnetic Fields Amplified by Nonlinear Turbulent Dynamo},
 volume = {975},
 ISSN = {1538-4357},
 url = {http://doi.org/10.3847/1538-4357/ad79fb},
 doi = {10.3847/1538-4357/ad79fb},
 number = {1},
 journal = {The Astrophysical Journal},
 publisher = {American Astronomical Society},
 author = {Zhang, Chao and Xu, Siyao},
 year = {2024},
 month = oct,
 pages = {65}
}

@article{Kuhlen2025b,
  title = {Diffusion of Relativistic Charged Particles and Field Lines in Isotropic Turbulence. II. Analytical Models},
  volume = {992},
  doi = {10.3847/1538-4357/adee94},
  number = {1},
  journal = {The Astrophysical Journal},
  publisher = {American Astronomical Society},
  author = {Kuhlen,  Marco and Mertsch,  Philipp and Phan,  Vo Hong Minh},
  year = {2025},
  month = oct,
  pages = {11}
}

@article{Kuhlen2025a,
  title = {Diffusion of Relativistic Charged Particles and Field Lines in Isotropic Turbulence. I. Numerical Simulations},
  volume = {992},
  doi = {10.3847/1538-4357/adee9a},
  number = {1},
  journal = {The Astrophysical Journal},
  author = {Kuhlen,  Marco and Mertsch,  Philipp and Phan,  Vo Hong Minh},
  year = {2025},
  month = oct,
  pages = {10}
}

@article{Ginzburg1966,
  title = {Origin of cosmic rays},
  volume = {9},
  doi = {10.1070/pu1966v009n02abeh002871},
  number = {2},
  journal = {Soviet Physics Uspekhi},
  author = {Ginzburg,  V. L. and Syrovatski\u{i},  S. I.},
  year = {1966},
  month = feb,
  pages = {223--235}
}

@article{Jokipii1966,
  title = {Cosmic-Ray Propagation. I. Charged Particles in a Random Magnetic Field},
  volume = {146},
  doi = {10.1086/148912},
  journal = {The Astrophysical Journal},
  author = {Jokipii,  J. R.},
  year = {1966},
  month = nov,
  pages = {480}
}

@article{Jokipii1967,
  title = "Cosmic-Ray Propagation. {II}. {D}iffusion in the Interplanetary Magnetic Field",
  volume = {149},
  doi = {10.1086/149265},
  journal = {The Astrophysical Journal},
  author = {Jokipii,  J. R.},
  year = {1967},
  month = aug,
  pages = {405}
}

@article{Forman1974,
  title = {Cosmic-Ray Streaming Perpendicular to the Mean Magnetic Field},
  volume = {192},
  doi = {10.1086/153087},
  journal = {The Astrophysical Journal},
  author = {Forman,  M. A. and Jokipii,  J. R. and Owens,  A. J.},
  year = {1974},
  month = sep,
  pages = {535---540}
}

@article{Giacalone1994,
  title = {Charged-particle motion in multidimensional magnetic-field turbulence},
  volume = {430},
  doi = {10.1086/187457},
  journal = {The Astrophysical Journal},
  author = {Giacalone,  J. and Jokipii,  J. R.},
  year = {1994},
  month = aug,
  pages = {L137}
}

@article{Reichherzer2025,
  title = {Efficient Micromirror Confinement of Sub-Teraelectronvolt Cosmic Rays in Galaxy Clusters},
  author = {Reichherzer, Patrick and Bott, Archie F. A. and Ewart, Robert J. and Gregori, Gianluca and Kempski, Philipp and Kunz, Matthew W. and Schekochihin, Alexander A.},
  year = {2025},
  journal = {Nature Astronomy},
  volume = {9},
  number = {3},
  pages = {438--448},
  doi = {10.1038/s41550-024-02442-1}
}

@article{Chhiber2021,
 title = {Magnetic field line random walk and solar energetic particle path lengths: Stochastic theory and {PSP/IS$\odot$IS} observations},
 volume = {650},
 doi = {10.1051/0004-6361/202039816},
 journal = {Astronomy \& Astrophysics},
 author = {Chhiber, R. and Matthaeus, W. H. and Cohen, C. M. S. and Ruffolo, D. and Sonsrettee, W. and Tooprakai, P. and Seripienlert, A. and Chuychai, P. and Usmanov, A. V. and Goldstein, M. L. and McComas, D. J. and Leske, R. A. and Szalay, J. R. and Joyce, C. J. and Cummings, A. C. and Roelof, E. C. and Christian, E. R. and Mewaldt, R. A. and Labrador, A. W. and Giacalone, J. and Schwadron, N. A. and Mitchell, D. G. and Hill, M. E. and Wiedenbeck, M. E. and McNutt, R. L. and Desai, M. I.},
 year = {2021},
 month = jun,
 pages = {A26}
}

@article{Lemoine2023,
 title = {Particle transport through localized interactions with sharp magnetic field bends in {MHD} turbulence},
 volume = {89},
 ISSN = {1469-7807},
 url = {http://doi.org/10.1017/S0022377823000946},
 doi = {10.1017/s0022377823000946},
 number = {5},
 journal = {Journal of Plasma Physics},
 publisher = {Cambridge University Press (CUP)},
 author = {Lemoine, Martin},
 year = {2023},
 month = sep,
 pages={175890501},
}

@book{Hasegawa1975,
  title = {Plasma Instabilities and Nonlinear Effects},
  author = {Hasegawa, Akira},
  year = {1975},
  publisher = {Springer},
  location = {Berlin},
  doi = {10.1007/978-3-642-65980-5},
}

@article{Pommois2001,
 title = {Field line diffusion in solar wind magnetic turbulence and energetic particle propagation across heliographic latitudes},
 volume = {106},
 ISSN = {0148-0227},
 url = {http://doi.org/10.1029/2001JA900050},
 doi = {10.1029/2001ja900050},
 number = {A11},
 journal = {Journal of Geophysical Research: Space Physics},
 publisher = {American Geophysical Union (AGU)},
 author = {Pommois, P. and Veltri, P. and Zimbardo, G.},
 year = {2001},
 month = nov,
 pages = {24965--24978}
}

@inproceedings{Boris1970,
  title={Relativistic plasma simulation-optimization of a hybrid code},
  author={Boris, Jay P.},
  booktitle={Proc. 4th Conf. Num. Sim. Plasmas},
  pages={3--67},
  year={1970}
}

@article{Weibel1959,
  doi = {10.1103/physrevlett.2.83},
  year = {1959},
  month = feb,
  volume = {2},
  number = {3},
  pages = {83--84},
  author = {Erich S. Weibel},
  title = {Spontaneously Growing Transverse Waves in a Plasma Due to an Anisotropic Velocity Distribution},
  journal = {Phys. Rev. Lett.}
}

@article{Arber2015,
  doi = {10.1088/0741-3335/57/11/113001},
  year = {2015},
  month = sep,
  volume = {57},
  number = {11},
  pages = {113001},
  author = {T. D. Arber and K. Bennett and C. S. Brady and A. Lawrence-Douglas and M. G. Ramsay and N. J. Sircombe and P. Gillies and R. G. Evans and H. Schmitz and A. R. Bell and C. P. Ridgers},
  title = {Contemporary particle-in-cell approach to laser-plasma modelling},
  journal = {Plasma Phys. Control. Fusion}
}

@article{Kocharovsky2016,
  title={Analytical theory of self-consistent current structures in a collisionless plasma},
  volume={59}, 
  url={http://doi.org/10.3367/UFNe.2016.08.037893},
  doi={10.3367/ufne.2016.08.037893},
  number={12},
  journal={Physics Uspekhi},
  author={Kocharovsky, V. V. and Kocharovsky, {\relax Vl}. V. and Martyanov, V. {\relax Yu}. and Tarasov, S. V.},
  year={2016},
  month=dec,
  pages={1165--1210}
}

@article{Kuznetsov2023,
  title = {Quasilinear Simulation of the Development of Weibel Turbulence in Anisotropic Collisionless Plasma},
  volume = {137},
  url = {http://doi.org/10.1134/S1063776123120099},
  doi = {10.1134/s1063776123120099},
  number = {6},
  journal = {Journal of Experimental and Theoretical Physics},
  author = {Kuznetsov, A. A. and Nechaev, A. A. and Garasev, M. A. and Kocharovsky, {\relax Vl}. V.},
  year = {2023},
  month = dec,
  pages = {966--985}
}

@inproceedings{Webb2001,
       author = {Webb, G. M. and Kota, J. and Zank, G. P. and Lu, J. Y.},
        title = {The {BGK} {B}oltzmann equation and anisotropic diffusion},
    booktitle = {International Cosmic Ray Conference},
         year = 2001,
       volume = {8},
        month = aug,
        pages = {3326--3329},
}

@article{Hasselmann1970,
  title = {A Note on the Parallel Diffusion Coefficient},
  volume = {162},
  doi = {10.1086/150736},
  journal = {The Astrophysical Journal},
  author = {Hasselmann,  K. and Wibberenz,  G.},
  year = {1970},
  month = dec,
  pages = {1049}
}

@article{Subedi2017,
  title = {Charged Particle Diffusion in Isotropic Random Magnetic Fields},
  volume = {837},
  doi = {10.3847/1538-4357/aa603a},
  number = {2},
  journal = {The Astrophysical Journal},
  publisher = {American Astronomical Society},
  author = {Subedi,  P. and Sonsrettee,  W. and Blasi,  P. and Ruffolo,  D. and Matthaeus,  W. H. and Montgomery,  D. and Chuychai,  P. and Dmitruk,  P. and Wan,  M. and Parashar,  T. N. and Chhiber,  R.},
  year = {2017},
  month = mar,
  pages = {140}
}

@article{Kubo1957,
  title = {Statistical-Mechanical Theory of Irreversible Processes. II. Response to Thermal Disturbance},
  volume = {12},
  doi = {10.1143/jpsj.12.1203},
  number = {11},
  journal = {Journal of the Physical Society of Japan},
  publisher = {Physical Society of Japan},
  author = {Kubo,  Ryogo and Yokota,  Mario and Nakajima,  Sadao},
  year = {1957},
  month = nov,
  pages = {1203--1211}
}

@BOOK{Davidson1972,
  author = {Davidson, R.~C.},
  title = "Methods in Nonlinear Plasma Theory",
  year = 1972,
  publisher = "Academic Press",
  address = "New York"
}

@article{Emelyanov2024,
  title = {Weibel Instability in the Presence of an External Magnetic Field: Analytical Results},
  volume = {66},
  doi = {10.1007/s11141-024-10326-7},
  number = {9},
  journal = {Radiophysics and Quantum Electronics},
  author = {Emelyanov,  N. A. and Kocharovsky,  Vl. V.},
  year = {2024},
  month = feb,
  pages = {664--678}
}

@article{Pokhotelov2012,
  title = {Weibel Instability in a Plasma with Nonzero External Magnetic Field},
  author = {Pokhotelov, O. A. and Balikhin, M. A.},
  year = {2012},
  journal = {Annales Geophysicae},
  volume = {30},
  number = {7},
  pages = {1051--1054},
  doi = {10.5194/angeo-30-1051-2012},
}

@article{Ibscher2012,
  title = "On the Existence of {W}eibel Instability in a Magnetized Plasma. {II}. Perpendicular Wave Propagation: The Ordinary Mode",
  author = {Ibscher, D. and Lazar, M. and Schlickeiser, R.},
  year = 2012,
  journal = {Physics of Plasmas},
  volume = {19},
  number = {7},
  pages = {072116},
  doi = {10.1063/1.4736992},
}

\clearpage

\begin{table}[htbp!]
\begin{tabular}{cccccc}
$\eta$ & $L_{\mathrm{cor}\parallel}/L_{\mathrm{cor}\perp}$ & $L_{\mathrm{cor}\parallel}\bar k_\parallel$ & $L_{\mathrm{cor}\parallel}\Delta k_\parallel$ & $L_{\mathrm{cor}\perp}\bar k_\perp$ & $L_{\mathrm{cor}\perp}\Delta k_\perp$ \\
0.04 & 1.34 & 12.24 & 13.04 & 10.13 & 8.09 \\
0.08 & 1.35 & 9.81 & 8.08 & 12.00 & 6.88 \\
0.16 & 1.37 & 5.66 & 4.46 & 17.92 & 8.74 \\
0.32 & 1.58 & 4.08 & 3.92 & 15.16 & 6.96 \\
0.50 & 1.64 & 3.80 & 3.98 & 6.30 & 3.03 \\
0.57 & 1.86 & 3.04 & 2.58 & 9.26 & 4.36 \\
0.80 & 2.84 & 3.15 & 3.38 & 2.97 & 1.35 \\
0.85 & 3.23 & 2.67 & 2.55 & 3.22 & 1.47 \\
0.96 & 5.17 & 1.98 & 1.78 & 1.87 & 0.87 \\
1.00 & 2.53 & 1.80 & 3.23 & 1.36 & 0.89 \\
\end{tabular}
\caption{Ratios of parallel ($L_{\mathrm{cor}\parallel}$) and perpendicular ($L_{\mathrm{cor}\perp}$) correlation lengths and normalized parallel and perpendicular mean wavenumbers ($\bar k_\parallel$, $\bar k_\perp$) and spectrum widths ($\Delta k_\parallel$, $\Delta k_\perp$) for different turbulence levels $\eta$ of the magnetic field}
\label{tbl:params}
\end{table}

\begin{table}[htbp!]
\begin{tabular}{ccccc}
$\eta$ & $\alpha_{\parallel 1}$ & $\alpha_{\parallel 2}$ & $\alpha_{\perp 1}$ & $\alpha_{\perp 2}$ \\
0.04 & $-0.77$ & $-0.28$ & $-0.98$ & $-1.57$ \\
0.08 & $-0.83$ &   0.65  & $-0.99$ & $-1.13$ \\
0.16 & $-0.89$ &   0.36  & $-1.03$ & $-0.94$ \\ 
0.32 & $-0.83$ &   0.92  & $-0.96$ & $-0.96$\\
0.50 & $-0.44$ &   1.06  & $-1.09$ & $-0.92$\\ 
0.57 & $-0.81$ &   0.65  & $-0.85$ & $-0.83$\\
0.80 & $-1.02$ &   1.03  & $-0.89$ & $-0.74$\\
0.85 & $-0.95$ &   0.91  & $-1.01$ & $-0.70$\\
0.96 & $-0.67$ &   0.63  & $-0.62$ & $-0.60$\\
1.00 & $-0.21$ &   1.12  & $-1.59$ & 1.09
\end{tabular}
\caption{Slope coefficients, or powers in a linear scale, for the curves from Fig.~\ref{fig:kappaRig} calculated for the first (index~1) and last (index~2) two points}
\label{tbl:dims}
\end{table}

\end{document}